\documentclass{article}
\usepackage[english]{babel}
\usepackage[letterpaper,top=2in,bottom=2in,left=2in,right=2in,marginparwidth=1.75cm]{geometry}
\usepackage{amsmath}
\usepackage{graphicx}
\usepackage{authblk}
\usepackage{hyperref}
\hypersetup{
    colorlinks=true,
    linkcolor=blue,
    filecolor=magenta,      
    urlcolor=blue,
}

\title{An Execution Fingerprint Dictionary \\ 
for HPC Application Recognition}

\author[1]{Thomas Jakobsche}
\author[2]{Nicolas Lachiche}
\author[1]{\\Aurélien Cavelan}
\author[1]{Florina M. Ciorba}

\affil[1]{Department of Mathematics and Computer Science, University of Basel, Switzerland}
\affil[2]{Data Science and Knowledge, University of Strasbourg, France}
\affil[ ]{Email: \textsuperscript{1}\{firstname.lastname\}@unibas.ch, \textsuperscript{2}nicolas.lachiche@unistra.fr}

\date{}

\begin{document}
\maketitle

\begin{abstract}
Applications running on HPC systems waste time and energy if they: (a)~use resources inefficiently, (b)~deviate from allocation purpose (e.g. cryptocurrency mining), or (c)~encounter errors and failures. It is important to know which applications are running on the system, how they use the system, and whether they have been executed before.
To recognize known applications during execution on a noisy system, we draw inspiration from the way Shazam recognizes known songs playing in a crowded bar.
Our contribution is an Execution Fingerprint Dictionary (EFD) that stores execution fingerprints of system metrics (keys) linked to application and input size information (values) as key-value pairs for application recognition.
Related work often relies on extensive system monitoring (many system metrics collected over large time windows) and employs machine learning methods to identify applications.
Our solution only uses the first 2 minutes and a single system metric to achieve F-scores above 95 percent, providing comparable results to related work but with a fraction of the necessary data and a straightforward mechanism of recognition. \\
\end{abstract}

\clearpage

\section{Introduction}
Scientific applications need high computing power provided by High Performance Computing (HPC) systems in order to answer scientific research questions. Operating HPC systems is costly and applications are competing for resources. It is therefore necessary to ensure orderly and efficient operations of HPC systems. The associated challenges are efficient energy and resource management, optimal resource allocation and scheduling, and orderly operations in the face of anomalies (including errors and misuse).

Operational Data Analytics (ODA) refers to operational data analysis for HPC system optimization~\cite{bourassa2019operational}. Monitoring and Operational Data Analytics (MODA) is a broader term that also encompasses data collection (monitoring).
MODA is a research field with the purpose of improving HPC operations and research. 
Through MODA, answers and actionable insights can be provided to the aforementioned challenges that go beyond system health checks. 
MODA also raises its own challenges: avoid additional overhead, avoid heavy storage requirements, and provide low-latency responses.

\textbf{Motivation:}
Applications running on HPC systems waste time and energy if they: (a)~use resources inefficiently, (b)~deviate from allocation purpose (e.g. cryptocurrency mining~\cite{crypto2020attack}), or (c)~encounter errors and failures. 
It is important to know which applications are running on the system, how they use the system, and whether they have been executed before.
Frequently used applications (e.g. for molecular dynamics or earthquake simulation) are executed repeatedly over time, potentially with different input sizes and by different users.
Currently there is no mechanism for keeping track of past application executions~\cite{yamamoto2018classifying}.

If we keep track of application executions and recognize that a job executes a known application, we can: 
(a)~make predictions about resource usage based on executions in the past (improving job scheduling~\cite{gaussier2015improving} and predicting energy consumption~\cite{shoukourian2014predicting}), 
(b)~detect deviations from past resource usage (indicating anomalies and potential errors), 
(c)~detect resource usage of known malicious applications (e.g. cryptocurrency mining~\cite{ates2018taxonomist}), and
(d)~lower power consumption by reducing CPU frequency for memory-bound applications~\cite{ott2020global}.

\textbf{Problem statement:}
To enable the aforementioned scenarios, we need to first find a mechanism to recognize the repeated execution of known applications in the presence of system noise and perturbations. 
In order to not interfere with the system itself, we want to \textit{understand how applications can be recognized in a lightweight and explainable manner}. 

\textbf{Existing solutions:}
Related work often relies on extensive system monitoring (many system metrics collected over large time windows) and employs machine learning methods to identify applications~\cite{taxonomist2018artifact}\cite{motaki2019gath}\cite{ramos2019accurate}\cite{liu2020characterization}. 
Such solutions are unattractive for use in production because they are too data-intensive and only provide high-latency responses after execution.

\textbf{Proposed solution:}
To recognize known applications during execution on a noisy system, we draw inspiration from the way Shazam recognizes known songs playing in a crowded bar~\cite{wang2006shazam}. 
Our contribution is an Execution Fingerprint Dictionary (EFD) that stores execution fingerprints of system metrics (keys) linked to application and input size information (values) as key-value pairs for application recognition.

\textbf{Evaluation:}
We evaluate the proposed EFD through experiments on a public data set containing labeled data (system metrics from CPU, memory, and network) collected during repeated executions of multiple applications with different input sizes on multiple nodes~\cite{taxonomist2018artifact}. 
Our solution only uses the first 2 minutes and a single system metric to achieve F-scores above 95 percent, providing comparable results to related work but with a fraction of the necessary data and a straightforward mechanism of recognition.

\section{Related work}
Taxonomist by Ates et al.~\cite{ates2018taxonomist} used a machine learning approach to classify applications executing on individual nodes. They used 721 system metrics and a time window spanning the whole execution. 
The authors mentioned but did not evaluate the potential to use their approach for smaller time windows.
For the EFD evaluation, we rely on the dataset~\cite{taxonomist2018artifact} used in the aforementioned work and compare against their results. 
Liu et al.~\cite{liu2020characterization} used performance counters to classify HPC applications with machine learning. They worked with application labels extracted from the executable name, which does not contain information about input sizes. 
Motaki et al.~\cite{motaki2019gath} used the Gath-Geva clustering algorithm on resource usage data to build clusters of similar applications. They discussed the potential of clusters for application identification but did not evaluate their solution from this perspective. 
Ramos et al.~\cite{ramos2019accurate} relied on performance counters to cluster similar applications. The authors noted that the same application is sometimes present in multiple clusters, caused by input sizes changing application characteristics or specific input sizes only triggering specific parts of the code, thus altering application behavior. 

Existing solutions often generalize application characteristics over different input sizes and node configurations. 
Generalization can be problematic if input sizes and node configuration change application behavior.
In contrast, we do not generalize characteristics and we do not use machine learning on large amounts of data. Inspired by Shazam, we build a dictionary for execution fingerprints linked to application information of past executions. 
Compared to related work, our approach uses considerably less data and only requires a small time interval at the beginning of an execution.

Yamamoto et al.~\cite{yamamoto2018classifying} followed an alternative approach: classification of applications based on static job information (e.g. job script, and executable files). 
Characteristics, like power consumption, can be estimated before execution based on applications with similar job information that were executed in the past. 
In contrast to their work, we do not generalize based on static job information, we directly recognize individual applications based on resource usage.

Recognizing application executions is similar to recognizing music songs. A well known music recognition service is Shazam, which recognizes short recordings of songs played in potentially noisy environments~\cite{wang2006shazam}. Shazam's recognition mechanism has 4 aspects that are relevant for this work: 
(1) using frequency as a statistical feature, 
(2) combining frequency peaks into fingerprints, 
(3) temporally aligning fingerprints for recognition, and 
(4) storing fingerprints in a hash-based lookup table~\cite{wang2003shazam}. 
Our work is a \textbf{\textit{proof of concept}} that focuses on the fourth aspect and uses a lookup table in the form of a dictionary for execution fingerprints of HPC applications. 
As statistical feature we compute the mean of resource usage measurements during a small time interval at the beginning of an execution. We (currently) do not employ combinatorial fingerprints or temporal alignment.

\section{Method}
\begin{figure}[h]
\centering
\includegraphics[width=8.7cm]{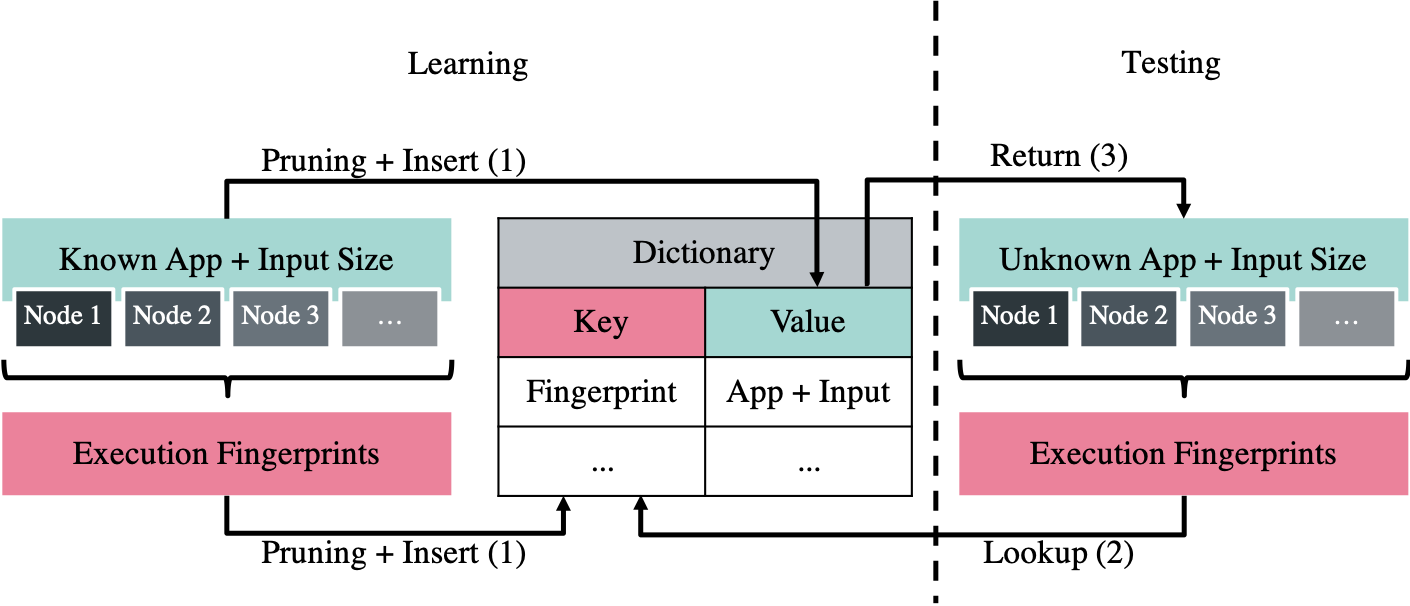}
\caption{Overview of the execution fingerprint dictionary-based application recognition mechanism. (1)~Per-node resource usage measurements are rounded to reduce precision (\textit{pruning}) and added to the dictionary with application information as key-value pairs.
(2)~Fingerprints of unlabeled executions are looked up. 
(3)~Application names are returned to the user if matches are found.}
\label{fig:overview}
\end{figure}

Our solution for application recognition is a dictionary-based approach that stores execution fingerprints linked to application information. Figure~\ref{fig:overview} provides an overview of the Execution Fingerprint Dictionary (EFD). 

\textbf{Learning:}
The dictionary is built by learning on a set of repeated executions with known application names.
We construct execution fingerprints (keys) that are linked to application information (values) and stored as key-value pairs in the dictionary. 
Fingerprints contain the rounded mean of measurements for individual nodes of a specific system metric during a particular time interval of an execution. Fingerprints consist of: (a)~metric name, (b)~node ID, (c)~time interval, and (d)~rounded mean.
An example fingerprint might look like this: [nr\_mapped\_vmstat, 0, [60:120], 6000.0].

An example EFD for the system metric \textit{nr\_mapped\_vmstat} is shown in Table~\ref{tab:example_dict}, where the 'key' column shows application execution fingerprints.
We chose the interval between 60 and 120 seconds after the beginning of an application execution to avoid the perturbations in the initialization phase while still reporting results relatively early during an execution.

\textbf{Pruning:}
Computing the mean produces precise floating point values that are unlikely to repeat due to system perturbations and noise. Computing distance measures for every example introduces unnecessary computational steps. Inspired by Shazam, we continue with low complexity by relying on dictionary-based matching of fingerprints with rounded values.

To get matching fingerprints, we need to guarantee that the same measurement gets rounded in the same way during training and testing. Therefore, we need to know how to round a measurement before seeing it. 
For this purpose we employ a rounding method with a configurable parameter called \textit{rounding depth}. The rounding mechanism is showcased in Table~\ref{tab:rounding}. \textit{Rounding depth} defines the position of a non-zero digit, counting from the left, to which we will round. 

\begin{table}[h]
\centering
\caption{Rounding Depth for Measurements}
\begin{tabular}{c|cccccc}
Original & \multicolumn{6}{c}{Rounding Depth}          \\
Value    & ... & 5 & 4      & 3      & 2      & 1      \\ \hline
1358.0   & ... & - & 1358.0 & 1360.0 & 1400.0 & 1000.0 \\
5.28     & ... & - & -      & 5.28   & 5.3    & 5.0    \\
0.038    & ... & - & -      & -      & 0.038  & 0.04
\end{tabular}
\label{tab:rounding}
\end{table}

Similar but distinct measurements will be rounded to the same fingerprint. 
Fingerprints (keys) are unique in our dictionary, rounding drastically reduces the number of entries and "\textit{prunes}" the dictionary. 
No \textit{pruning} will lead to precise fingerprints that have high exclusiveness and low repetition count. 
Excessive \textit{pruning} will lead to heavily rounded fingerprints that have low exclusiveness and high repetition count. 
\textit{Rounding depth} is the only tunable parameter in the EFD. During the learning phase we find the optimal \textit{rounding depth} through cross-fold validation within the training set.

\textbf{Testing:}
Given an execution of the test set we construct fingerprints with the same \textit{rounding depth} as in the learning phase, for each node of the execution. Fingerprints of each node are looked up in the dictionary, and the most matched application name is returned. If multiple applications have the same number of matches (potentially caused by key collisions) the EFD cannot distinguish between them and will return an array of these application names. For evaluation purpose we consider the first application name in the array.

\section{Evaluation and Results}
To evaluate the EFD we compare F-scores (harmonic mean of precision and recall) to the Taxonomist by Ates et al.~\cite{ates2018taxonomist}. 
The \textit{application name}--\textit{input size} pairs contained in the dataset are shown in Table~\ref{tab:dataset}. 
The dataset contains repeated executions on multiple nodes. 
The publicized dataset is a subset of the original, containing only one third of the repeated executions and only 562 of the original 721 system metrics. 
System metrics from CPU, memory, and network were collected every second from every node of a given execution through the LDMS monitoring solution~\cite{agelastos2014lightweight}. For more information on the applications, the HPC system, and the employed monitoring solution, we refer to the work of Ates et al.~\cite{ates2018taxonomist}.

\begin{table}[h]
\centering
\caption{Dataset used for Evaluation}
\begin{tabular}{c|c|c|c}
Applications & \begin{tabular}[c]{@{}c@{}}Input\\ Sizes\end{tabular} & \begin{tabular}[c]{@{}c@{}}Node\\ Count\end{tabular} & \begin{tabular}[c]{@{}c@{}}Repeated\\ Executions\end{tabular} \\ \hline
\begin{tabular}[c]{@{}c@{}}FT, MG, SP, LU, BT, CG, \\ CoMD, miniGhost*, \\ miniAMR*, miniMD*, kripke*\end{tabular} & \begin{tabular}[c]{@{}c@{}}X, Y, Z\\ L*\end{tabular}  & \begin{tabular}[c]{@{}c@{}}4\\ 32\end{tabular}       & \begin{tabular}[c]{@{}c@{}}30\\ 6\end{tabular}                \\ \hline
\multicolumn{4}{c}{* Input L is only available for a subset of applications.}
\end{tabular}
\label{tab:dataset}
\end{table}

\textbf{Experiments:} We evaluate the EFD through different offline experiments that showcase utility and recognition capabilities. Executions have two identifying dimensions: application name and input size. The experiments differ in the way the learning and testing sets are split along those dimensions:
\begin{enumerate}
    \item Normal fold: 5-fold cross validation on the full dataset (all applications and inputs in both learning and testing.)
    \item Soft input: Extends normal fold, removes individual input sizes from learning, testing sets stay the same.
    \item Soft unknown: Extends normal fold, removes individual applications from learning, testing sets stay the same.
    \item Hard input: Learning set contains 3 out of 4 input sizes and testing set only the 4th (exclusively testing with unknown input sizes). 
    \item Hard unknown: Learning set contains 10 out of 11 applications and testing set only the 11th (exclusively testing with unknown applications). 
\end{enumerate}

F-score and cross-fold validation are implemented using the sci-kit learn library~\cite{pedregosa2011scikit}. 
In the "soft / hard input" experiments, we test the capability of the dictionary to recognize applications with unknown input sizes. 
Each input size is removed once and results are averaged. 
We look at the application name to decide correctness (e.g. returning FT\_X for FT\_Y is considered correct). 
In the "soft / hard unknown" experiments we test whether the dictionary wrongfully recognizes unknown applications. 
In the latter case we consider finding no matching fingerprints as a correct prediction for unknown applications. 
Each application is removed once and results are averaged. 

\begin{figure}[h]
\centering
\includegraphics[width=7cm]{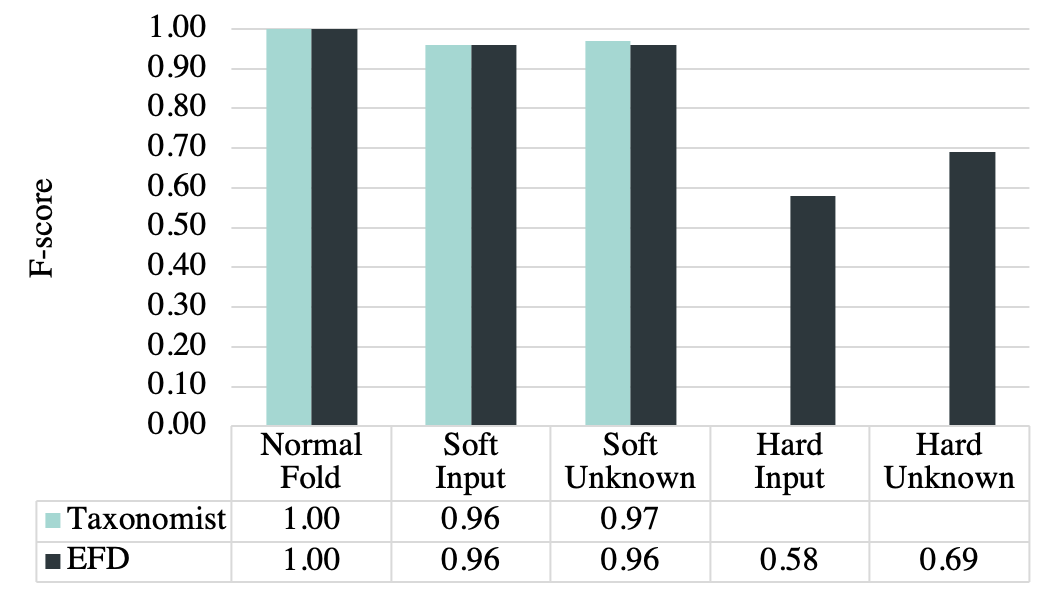}
\caption{Comparison between Taxonomist (using 721 system metrics and the entire execution time window) and EFD (using only 1 system metric \textit{nr\_mapped\_vmstat} and only the first 2 minutes of the execution time window). The 'hard input' and 'hard unknown' experiments were not conducted in the Taxonomist~\cite{ates2018taxonomist}.}
\label{fig:results}
\end{figure}

"Soft" experiments reflect normal operations, where some applications and inputs are known, while others are unknown. "Hard" experiments have very harsh criteria for success, because we exclusively test with unknown applications or input sizes.
Figure~\ref{fig:results} shows the results of our solution compared to the results reported by the Taxonomist. 
Table~\ref{tab:system_metrics} shows an excerpt of individual results of other system metrics using the normal fold experiment.

\begin{table}[h]
\centering
\caption{Excerpt of Individual System Metric Results}
\begin{tabular}{l|c}
\multicolumn{1}{c|}{System Metric Name} & \begin{tabular}[c]{@{}c@{}}F-score\\ Normal Fold\end{tabular} \\ \hline
nr\_mapped\_vmstat              & 1.0       \\ \hline
Committed\_AS\_meminfo          & 1.0       \\ \hline
nr\_active\_anon\_vmstat        & 1.0       \\ \hline
nr\_anon\_pages\_vmstat         & 1.0       \\ \hline
Active\_meminfo                 & 0.99      \\ \hline
Mapped\_meminfo                 & 0.99      \\ \hline
AnonPages\_meminfo              & 0.97      \\ \hline
MemFree\_meminfo                & 0.97      \\ \hline
PageTables\_meminfo             & 0.97      \\ \hline
nr\_page\_table\_pages\_vmstat  & 0.97      \\ \hline 
AMO\_PKTS\_metric\_set\_nic     & 0.96      \\ \hline 
AMO\_FLITS\_metric\_set\_nic    & 0.95      \\ \hline 
PI\_PKTS\_metric\_set\_nic      & 0.95      \\ \hline 
...                             & ...
\label{tab:system_metrics}
\end{tabular}
\end{table}

\section{Discussion}
An example EFD based on the system metric \textit{nr\_mapped\_vmstat} is shown in Table~\ref{tab:example_dict} and discussed below. This EFD was built with a subset of the applications and input sizes and a fixed \textit{rounding depth} to reduce the size and show a full example.

\textbf{Exclusive fingerprints and collisions:}
The example EFD in Table~\ref{tab:example_dict} shows how recognition can be achieved through application exclusive execution fingerprints. It also shows a collision between SP and BT from the NAS parallel benchmark suite~\cite{bailey1991parallel}, which are applications known to be similar~\cite{ma2009approach}. 
If multiple applications have the same number of matches, the EFD cannot distinguish between them and will return an array of these application names. 
For evaluation purpose we consider the first application name in the array, in this case SP. The example EFD was fixed to \textit{rounding depth} 2.
\textit{Rounding depth} 3 avoids this collision and also recognizes BT.

\textbf{Recognizing applications with unknown input:}
Depending on the application and system metric considered, execution fingerprints repeat even for different application input sizes. 
This, however, does not apply to all applications (e.g. miniAMR). 
This behavior can be exploited to recognize some applications executed with unknown input sizes. 
HPC workload variation is problematic for our recognition approach and the question whether we can find suitable input independent system metrics for larger sets of applications remains open. The results of the "hard input" experiment, shown in Figure~\ref{fig:results}, show that there is room for improvement.

\begin{table}[h] \tiny \centering
\caption{Example Execution Fingerprint Dictionary}
\begin{tabular}{c|c|c|c|c}
\multicolumn{4}{c|}{Key (execution fingerprint)}                                 & Value      \\ \hline
Metric Name        & Node & Interval     & Mean    & Application + Input Size                 \\ \hline
nr\_mapped\_vmstat & 0    & {[}60:120{]} & 6000.0  & ft\_X, ft\_Y, ft\_Z                      \\ \hline
nr\_mapped\_vmstat & 1    & {[}60:120{]} & 6000.0  & ft\_X, ft\_Y, ft\_Z                      \\ \hline
nr\_mapped\_vmstat & 2    & {[}60:120{]} & 6000.0  & ft\_X, ft\_Y, ft\_Z                      \\ \hline
nr\_mapped\_vmstat & 3    & {[}60:120{]} & 6000.0  & ft\_X, ft\_Y, ft\_Z                      \\ \hline
nr\_mapped\_vmstat & 0    & {[}60:120{]} & 6100.0  & mg\_X, mg\_Y, mg\_Z                      \\ \hline
nr\_mapped\_vmstat & 1    & {[}60:120{]} & 6100.0  & mg\_X, mg\_Y, mg\_Z                      \\ \hline
nr\_mapped\_vmstat & 2    & {[}60:120{]} & 6100.0  & mg\_X, mg\_Y, mg\_Z                      \\ \hline
nr\_mapped\_vmstat & 3    & {[}60:120{]} & 6100.0  & mg\_X, mg\_Y, mg\_Z                      \\ \hline
nr\_mapped\_vmstat & 0    & {[}60:120{]} & 7600.0  & sp\_X, sp\_Y, sp\_Z, bt\_X, bt\_Y, bt\_Z \\ \hline
nr\_mapped\_vmstat & 1    & {[}60:120{]} & 7500.0  & sp\_X, sp\_Y, sp\_Z, bt\_X, bt\_Y, bt\_Z \\ \hline
nr\_mapped\_vmstat & 2    & {[}60:120{]} & 7500.0  & sp\_X, sp\_Y, sp\_Z, bt\_X, bt\_Y, bt\_Z \\ \hline
nr\_mapped\_vmstat & 3    & {[}60:120{]} & 7100.0  & sp\_X, sp\_Y, sp\_Z, bt\_X, bt\_Y, bt\_Z \\ \hline
nr\_mapped\_vmstat & 0    & {[}60:120{]} & 8400.0  & lu\_X, lu\_Y, lu\_Z                      \\ \hline
nr\_mapped\_vmstat & 1    & {[}60:120{]} & 8300.0  & lu\_X, lu\_Y, lu\_Z                      \\ \hline
nr\_mapped\_vmstat & 2    & {[}60:120{]} & 8300.0  & lu\_X, lu\_Y, lu\_Z                      \\ \hline
nr\_mapped\_vmstat & 3    & {[}60:120{]} & 8300.0  & lu\_X, lu\_Y, lu\_Z                      \\ \hline
nr\_mapped\_vmstat & 0    & {[}60:120{]} & 7900.0  & miniGhost\_X, miniGhost\_Y, miniGhost\_Z \\ \hline
nr\_mapped\_vmstat & 1    & {[}60:120{]} & 7900.0  & miniGhost\_X, miniGhost\_Y, miniGhost\_Z \\ \hline
nr\_mapped\_vmstat & 2    & {[}60:120{]} & 7900.0  & miniGhost\_X, miniGhost\_Y, miniGhost\_Z \\ \hline
nr\_mapped\_vmstat & 3    & {[}60:120{]} & 7900.0  & miniGhost\_X, miniGhost\_Y, miniGhost\_Z \\ \hline
nr\_mapped\_vmstat & 0    & {[}60:120{]} & 7800.0  & miniAMR\_X                               \\ \hline
nr\_mapped\_vmstat & 1    & {[}60:120{]} & 7800.0  & miniAMR\_X                               \\ \hline
nr\_mapped\_vmstat & 2    & {[}60:120{]} & 7800.0  & miniAMR\_X                               \\ \hline
nr\_mapped\_vmstat & 3    & {[}60:120{]} & 7800.0  & miniAMR\_X                               \\ \hline
nr\_mapped\_vmstat & 0    & {[}60:120{]} & 8000.0  & miniAMR\_Y                               \\ \hline
nr\_mapped\_vmstat & 1    & {[}60:120{]} & 8000.0  & miniAMR\_Y                               \\ \hline
nr\_mapped\_vmstat & 2    & {[}60:120{]} & 8000.0  & miniAMR\_Y                               \\ \hline
nr\_mapped\_vmstat & 3    & {[}60:120{]} & 8000.0  & miniAMR\_Y                               \\ \hline
nr\_mapped\_vmstat & 0    & {[}60:120{]} & 11000.0 & miniAMR\_Z                               \\ \hline
nr\_mapped\_vmstat & 1    & {[}60:120{]} & 11000.0 & miniAMR\_Z                               \\ \hline
nr\_mapped\_vmstat & 2    & {[}60:120{]} & 11000.0 & miniAMR\_Z                               \\ \hline
nr\_mapped\_vmstat & 3    & {[}60:120{]} & 11000.0 & miniAMR\_Z                               \\ \hline
nr\_mapped\_vmstat & 2    & {[}60:120{]} & 10000.0 & miniAMR\_Z                               \\ \hline
nr\_mapped\_vmstat & 1    & {[}60:120{]} & 10000.0 & miniAMR\_Z                                                           
\end{tabular}
\label{tab:example_dict}
\end{table}

\textbf{The impact of node configuration:}
Certain applications use nodes in consistently different ways, e.g., SP and BT from the NAS parallel benchmark suite~\cite{bailey1991parallel}. The absence of fingerprints where the nodes are not used differently, indicate that all training examples used the nodes in this way. The Taxonomist evaluates and labels individual nodes, whereas the EFD evaluates the entire execution. Applications on HPC systems are executed on multiple that can be used differently. It, therefore, stands to reason that we recognize an application through all involved nodes.

\textbf{Measurement variation and system noise:}
Some \textit{application name}--\textit{input size} pairs produce more than one fingerprint per node (e.g. miniAMR\_Z). Different learning examples may produce different fingerprints due to measurement variations, potentially caused by system perturbations and noise. \textit{Pruning} the dictionary can soften the impact of measurement variations, but excessive \textit{pruning} will produce generic fingerprints that are no longer exclusive.

\textbf{Robustness against unknown applications:}
If unknown applications produce execution fingerprints that are not in the dictionary, they will not be recognized and thus correctly labeled as unknown. This is an in-built safeguard against unknown applications. However, it is unclear whether we can build exclusive fingerprints for larger sets of applications. Going forward, we can make fingerprints more exclusive by combining multiple system metrics and / or multiple time intervals from the execution time window. The results of the "hard unknown" experiment, shown in Figure~\ref{fig:results}, reflect the room for improvement regarding EFD's recognition robustness against unknown applications.

\section{Conclusion}
This work contributes a proof of concept that a Shazam-inspired dictionary-based approach can recognize repeated executions of applications on HPC systems. 
The proposed EFD solution recognizes a repeated application execution with F-scores above 95 percent within the first 2 minutes by only using a single system metric. 
These results provide comparable \mbox{F-scores} to related work but with a fraction of the necessary data and a straightforward mechanism of recognition.

The way application execution fingerprints are built allows the co-existence of fingerprints for different system metrics and time intervals within the same dictionary. 
This opens future work on more exclusive, temporally aligned, and combinatorial fingerprints, which would bring the EFD closer to the mechanism used by Shazam~\cite{wang2003shazam}. 
If application execution fingerprints are sufficiently exclusive, learning new applications is as simple as adding new keys to the dictionary. 
Populating the dictionary with different time intervals could enable resource usage prediction, by using the dictionary in reverse, 
namely by looking up applications to report potential future resource usage based on resource usage in the past.

\end{document}